\documentclass{appolb}
\usepackage{graphicx}
\usepackage{amsmath}
\usepackage{cite}
\usepackage{color}
\graphicspath{{figures/}}
\usepackage[utf8]{inputenc}
\usepackage[english]{babel}

\begin{document}
\title{The role of emotional variables in the classification and prediction of collective social dynamics
}

\author{Jan Chołoniewski\footnote{e-mail: choloniewski@if.pw.edu.pl} and Julian Sienkiewicz
\address{Physics in Economy and Social Sciences Research Group,\\ Faculty of Physics, Warsaw University of Technology,\\ Koszykowa 75, PL-00662 Warszawa, Poland}
\\ \vspace{10pt}
{
Janusz Hołyst
\address{Physics in Economy and Social Sciences Research Group,\\ Faculty of Physics, Warsaw University of Technology,\\ Koszykowa 75, PL-00662 Warszawa, Poland}
\address{ITMO University, \\19, Kronverkskiy av., 197101 Saint Petersburg, Russia}
}
\\ \vspace{10pt}
{Mike Thelwall\\
\address{Statistical Cybermetrics Research Group,\\ School of Mathematics and Computer Science, University of Wolverhampton,\\ Wulfruna Street, Wolverhampton WV1 1LY, United Kingdom}}
}
\maketitle
\begin{abstract}
We demonstrate the power of data mining techniques for the analysis of collective social dynamics within British Tweets during the Olympic Games 2012. The classification accuracy of online activities related to the successes of British athletes significantly improved when emotional components of tweets were taken into account, but employing emotional variables for activity prediction decreased the classifiers' quality. The approach could be easily adopted for any prediction or classification study with a set of problem-specific variables.
\end{abstract}

\section{Introduction}

Knowledge of human collective behaviour is a powerful tool in the hands of those that know how to use it. Advances in economics and social sciences can be insightful but quantitative descriptions are often absent. The reason might be the high complexity of many social systems, their highly nonlinear dynamics and internal relationships. Physicists have encountered these kinds of problems many times and have provided valuable conclusions -- if one knows only a little about the underlying microscopic system dynamics (such as a phase space, short- and long-distance interactions), then one can focus on the macroscopic outputs of the system using statistical methods \cite{prof1, prof2, prof3, prof4, urban, kutner, drozdz}. 

Following mass adoption of the internet, it became clear that gathering massive amounts of data about a people's behavior (on-line stock markets, social media) -- nearly impossible before -- had become easy. It is possible to track the activities of an entity but describing a person would be a great challenge for the scientists of the future. Nevertheless, internet data is extremely useful to study collective human behaviors. This approach has already been tested within some recent studies \cite{collective1,collective2,collective3}. Moreover, it provides quantitative methods for social dynamics and emotion analyses \cite{prof5, prof6, holyst, bbc}. The last cited model describes possible proceedings of online discussions taking into account the emotional context of posts -- e.g., if the first post in a discussion has strong positive positive emotions, what tends to happen to the rest of the discussion?

These new possibilities always create new challenges – including how to effectively manage the necessary big data.

\subsection{Data mining in science}
Astronomy and astrophysics had to deal with a data abundance since an invention of an automatic optical telescope and digital photograpy. An amount of highly magnified high resolution pictures of the sky heavily outnumbered a size of the labor force. 

Traditional methods to analyse bid data are too time-consuming -- even an army of astronomers would need to study such big datasets for years. This problem led to the development of statistical data exploration methods \cite{astro1}. For example, extracting the features of an object from a photo and applying a classification algorithm trained on empirical data enables the detection of various kinds of stars and galaxies. Data mining techniques have also been employed in large CERN projects -- e.g. in particle classifications conducted during the Higgs boson search experiments \cite{higgs}, but also in many other physical sciences – for example predicting certain kinds of crystals \cite{cristal} or graphen-like structures \cite{grafen} in solid-state physics.

Of course, data mining applications are not limited to physics and science in general (at the moment, these are not even their main areas). There are also commercial (eg. basket analyses \cite{basket}, customer behaviour research \cite{customer}), medical (eg. automated diagnostics \cite{medic1}) and economics (eg. credit scoring \cite{credit}) uses. The development of data mining techniques is inevitable and will greatly impact on our everyday life.

\subsection{Science in data mining}
Data mining perhaps began in the 1700s when Bayes presented his famous theorem about a conditional probability to the Royal Society but the real advances happened in the 20th century -- cluster analysis \cite{cluster}, genetic algorithms \cite{ga}, classification trees \cite{trees}, support vector machines \cite{svm} etc. Nowadays, these are also part of the new discipline of big data science. While most findings should be considered as being contributions from mathematics and computer science, data mining has bidirectional interactions with physics.

Physics-based approaches have led to new interesting new methods, such as a clustering algorithm based on quantum mechanics \cite{qdm}, support vector machine studies based on statistical physics \cite{svm_phys} or a cluster analysis method inspired by theoretical spin glasses research \cite{spin}.

In the face of the great benefits and potentials of data mining, it seems that developing new methods and improving existing ones could be a great help for the next generation of physicists.

\subsection{Main focus}
The main aims of the paper are to check (i) if one can predict online social dynamics using data mining techniques, (ii) if one can detect special events taking place in the real world from observations of online discussions, (iii) if detecting emotional content can contribute to a better performance for some social data mining techniques.

Twitter is a social website that reacts to some kinds of external impacts (i.e., medal winning during the 2012 Olympic Games in London). In this paper, its activities were predicted and its dynamics were classified. 
The relationships between the number of dimensions available, the methods applied, emotional data usage and the accuracy of trained classifiers were also analyzed. 

Note that previous statistical Twitter analyses have also yielded interesting results, such as stock market predictions \cite{twitter1} or extreme social-media events predictions \cite{js1}.
\section{Data}
\subsection{Original dataset}
The original dataset was gathered by the Statistical Cybermetrics Research Group from University of Wolverhampton and shared as part of the EU funded CyberEmotions project.

From 2012-07-11 to 2012-08-13 (2012 Olympic Games in London and about two weeks before), all tweets written in English and sent within a radius of 50 km around London were gathered if they contained a sports-related hashtag (eg. \textit{\#olympics}, \textit{\#100mrun}). The information about each tweet consists of its unique ID, text (up to 160 characters, including hashtags), date and time sent, author's user name, if a tweet was a response to another tweet -- ID and author's user name of that tweet, positive sentiment strength (scored by the SentiStrength classifier~\cite{senti}; from 1 to 5, where 1 -- no positive sentiment and 5 -- very strong positive sentiment), negative emotion strength (scored by the SentiStrength classifier; from -1 to -5, where -1 -- no negative sentiment and -5 -- very strong negative sentiment).

Tweets with no sentiment (i.e., a positive sentiment score of 1 and a negative sentiment score of -1) were discarded from the analysis.

\subsection{Aggregated and derived data}
The original dataset was aggregated into 3134 non-overlapping 15-minute time windows. In each window, features were calculated as presented in Tab.~\ref{tab:t1}. These variables were used as predictors for benchmark problems.

Peaks of activity and sentiment in the Twitter time series were detected using the algorithm described in \cite{billhauer} and the presence of a sentiment peak was marked as a feature of each time window (\textbf{sentiment peak -- SENT PEAK*}). The presence of an activity peak was later used as a benchmark for problem 3. 
\subsection{Medal data}
For some analyses, dates and times of British athletes winning a medal were recorded. Medal colors and sports events were also gathered but not used in the study. The complete list can be found in Appendix A.

\begin{table}[]\caption{Features in 15-minute windows used in the classification aggregated and derived from original dataset; features marked with star (*) are considered \textit{sentiment} dimensions in the study.}
\centering\noindent\makebox[\textwidth]{
\begin{tabular}{|p{.18\textwidth}|p{.32\textwidth}|p{.5\textwidth}|}
\hline
\textbf{Abbreviation} & \textbf{Description} & \textbf{How it was calculated?} \\ \hline
\hline
\textbf{ACT} & Activity & A number of tweets sent in a given 15-minute window \\ \hline
\textbf{SENT*} & Sentiment & A mean difference between a positive and a negative emotions score \\ \hline
\textbf{PERC UNIQ} & Percent of unique users & A number of unique users (tweet authors) divided by the number of tweets  \\ \hline
\textbf{PERC REP} & Percent of reply tweets & A number of tweet that are replies divided by the number of tweets  \\ \hline
\textbf{MEAN PL} & Post length & A mean tweet number of characters \\ \hline
\textbf{dACT} & First derivative of activity & A difference between ACT in the time windows and ACT in the previous time window \\ \hline
\textbf{dSENT*} & First derivative of sentiment & A difference between SENT in the time windows and SENT in the previous time window \\ \hline
\textbf{d2ACT} & Second derivative of activity & A difference between dACT in the time windows and dACT in the previous time window \\ \hline
\textbf{d2SENT*} & Second derivative of sentiment & A difference between dSENT in the time windows and dACT in the previous time window \\ \hline
\textbf{SENT PEAK*} & Sentiment peak & \textit{description in 2.2.}  \\ \hline
\end{tabular}}
\label{tab:t1}
\end{table}

\section{Benchmark problems}
Classifiers (i.e., data mining methods trained on a set of observations of given dimensionality) were provided information about the features in previous time windows and were tested on three two-class classification problems. A summary of class populations for each problem is in Tab.~\ref{tab:t2}.

The combination of features listed in Tab.~\ref{tab:t1} was considered in each time window as an observation vector.

Below we describe the tasks that the classifiers had to solve.
\subsection{Trend prediction}
\label{bp1}
Each time window was considered as an observation. The classifier had to determine if the activity component in the next time window would be lesser (class 1) or greater (class 2) than the activity in the present time window, knowing the present and the past feature vector.

\subsection{Threshold forecasting}
\label{bp2}
The procedure was as in~\ref{bp1}, but the classifier had to determine if the activity in a given 15-minute window would be lesser (class 1) or greater (class 2) than 500 tweets.

\subsection{Peak classification}
\label{bp3}
Each time window where an activity peak was detected was considered as an observation. The classifier had to determine if the peak co-occurred with an event in which a British athlete(s) had won a medal.
\begin{table}[]
\caption{Class counts for each benchmark problem}
\centering\noindent\makebox[\textwidth]{
\begin{tabular}{|l|l|l|l|}
\hline
\textbf{Problem} & \textbf{Observations} & \textbf{Class 1 count} & \textbf{Class 2 count} \\ \hline
\hline
Trend prediction & 3140 & 1614 (51.4\%) & 1516 (48.6\%) \\ \hline
Threshold exceed & 3140 & 1768 (56.3\%) & 1362 (43.7\%) \\ \hline
Peak classification & 516 & 482 (93.4\%) & 34 (7.6\%) \\ \hline
\end{tabular}}
\label{tab:t2}
\end{table}
\section{Methodology}
9 data mining methods were chosen for testing along with 17 features for the first and the second problem and 15 for the third problem. The accuracy for each pair ``method-set of features" (sets of features constructed from up to 6 elements) was tested on the benchmark problems.

All 2-, 3-, 4-, 5- and 6-element permutations of features were employed to train each method on a randomly chosen 80\% of the observations and tested on the remaining 20\% of the observations. The procedure was repeated 20 times for each pair.

\subsection{Data mining methods}
Applied methods:
\begin{itemize}
\item linear discriminant analysis \textbf{(LDA)},
\item naive Bayes \textbf{(NB)},
\item quadratic discriminant analysis \textbf{(QDA)},
\item regression tree \textbf{(REG TREE)},
\item supporting vector machines \textbf{(SVM)} with various cores;
\end{itemize}
SVM cores:
\begin{itemize}
\item linear \textbf{(SVM LIN)},
\item tangential \textbf{(SVM MLP)},
\item polynomial (3rd order) \textbf{(SVM POLY3)},
\item quadratic \textbf{(SVM QUAD)},
\item radial-based function \textbf{(SVM RBF)};
\end{itemize}

\subsection{Variables sets}
For each observation in the set, the following features were selected ($t$ means ``in the time window where the peak occurred", $t-1$ means ``in the previous time window" and so on) for first two problems (abbreviations described in Tab.~\ref{tab:t1}): ACT(t-1), ACT(t-2), ACT(t-3), SENT(t-1), SENT(t-2), SENT(t-3), dACT(t-1), dACT(t-2), dSENT(t-1), dSENT(t-2), d2ACT(t-1), d2ACT(t-2), d2SENT(t-1), d2SENT(t-2), PROC REP(t-1), PROC UNIQ(t-1), MEAN PL(t-1). For the third problem: ACT(t), ACT(t-1), ACT(t-2), SENT(t), SENT(t-1), SENT(t-2), SENT PEAK(t), SENT PEAK(t-1), SENT PEAK(t-2), PERC UNIQ(t), PERC UNIQ(t-1), PERC UNIQ(t-2), MEAN PL(t), MEAN PL(t-1), MEAN PL(t-2).

\section{Results}
For each benchmark problem, the relationships between the number of dimensions and classifier accuracy were analysed as well as the impact of applying a specific number of emotional dimensions.

\subsection{Trend prediction}
\begin{figure}[htb]
\centerline{
\includegraphics[width=12cm]{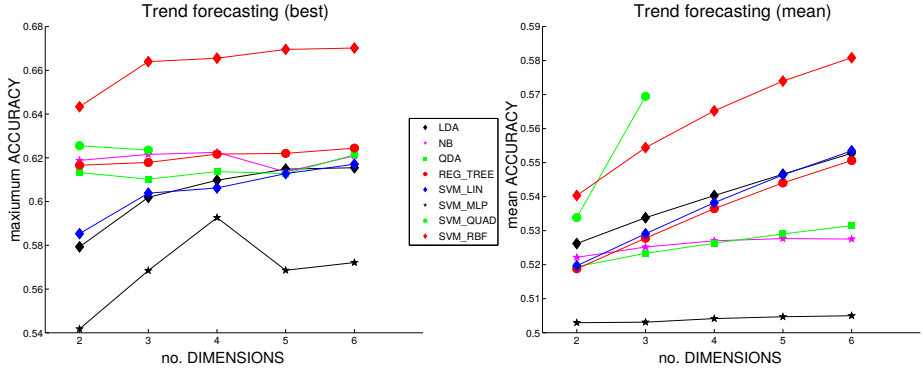}}
\caption{Maximum (left graph) and mean (right) accuracy of classifiers as a function of a number of variables (dimensions); standard deviations were not plotted for the sake of readability, they are of the order of $0.02$}
\label{fig:tp1}
\end{figure}

In Fig.~\ref{fig:tp1} the best (left) and the mean (right) accuracy of trend predicting (see subsection~\ref{bp1}) classifiers with a given dimensionality is presented. The best results are achieved for support vector machines with a radial-basis function as their core. These classifiers achieved an accuracy of 66\%. It is interesting that adding dimensions does not necessarily enhance the quality of the best classifiers. The \textbf{mean accuracy} of a classifier tends to increase with every additional dimension, however.

\begin{figure}[htb]
\centerline{
\includegraphics[width=12cm]{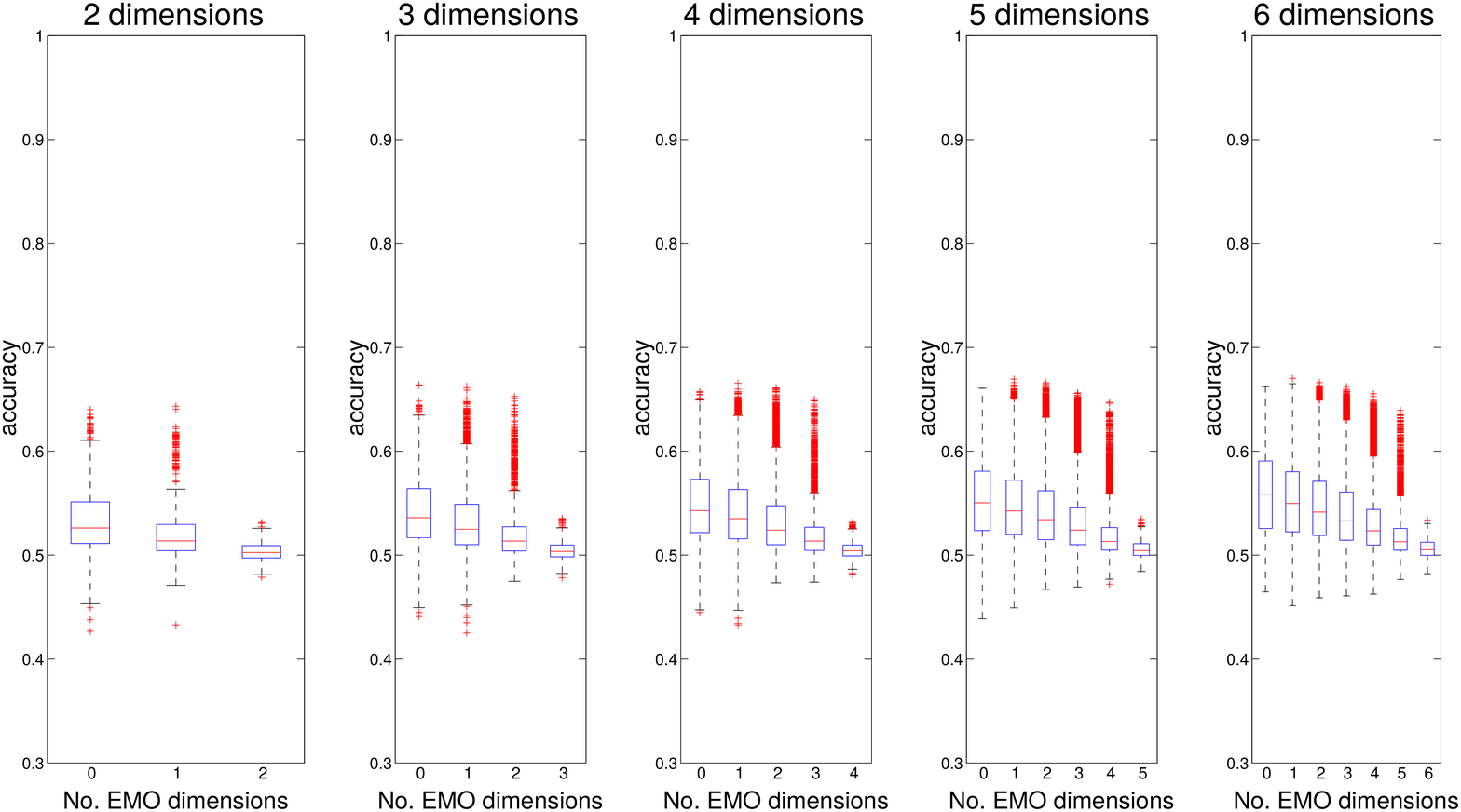}}
\caption{Box plot of \textbf{accuracy} for different numbers of variables used by \textbf{various classifiers} as a function of number of emotion-related variables;}
\label{fig:tp2}
\end{figure}

\begin{figure}[htb]
\centerline{
\includegraphics[width=12cm]{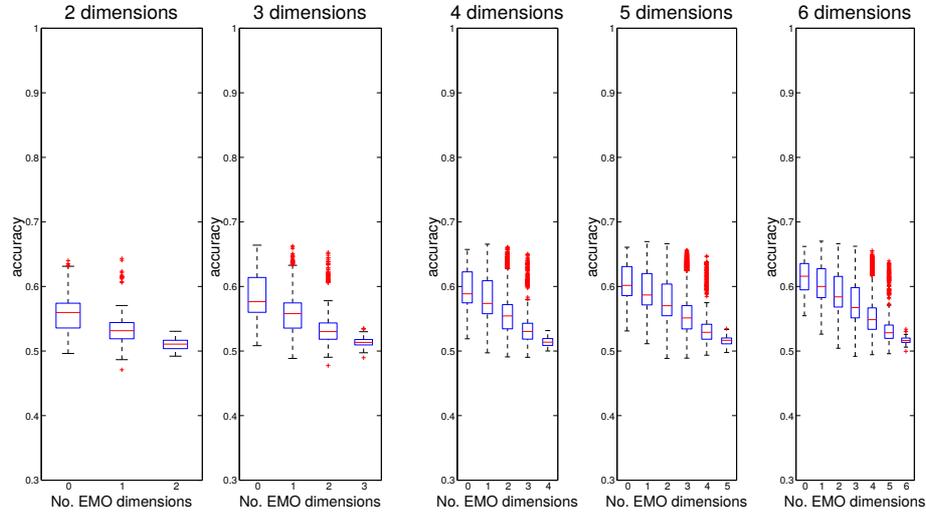}}
\caption{Box plot of \textbf{accuracy} for different numbers of variables used by \textbf{SVM with a radial basis core} as a function of number of emotion-related variables;}
\label{fig:tp3}
\end{figure}

Fig.~\ref{fig:tp2} shows the relationship between the accuracy of the trend predicting classifiers and the number of emotional dimensions for all methods, while in Fig.~\ref{fig:tp3} -- for support vector machines with radial basis core (the most accurate classifiers for the problem). It is clear that in this case using emotional dimensions does not enhance classification. Emotional features seem to carry no valuable information for this kind of exercises -- comparing accuracies for 2 dimensional classifiers with 0 emotional features (2-0emo) with 3-1emo, 4-2emo, 5-3emo and 6-4emo gives nearly the same result.

\subsection{Threshold exceed forecasting}
\begin{figure}[htb]
\centerline{
\includegraphics[width=12cm]{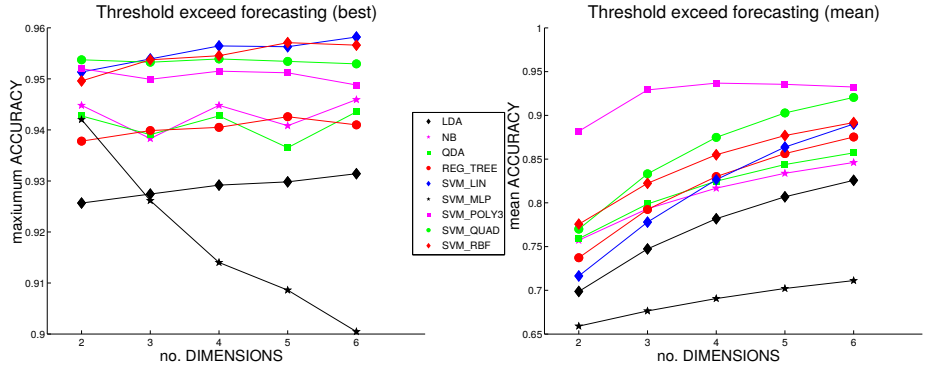}}
\caption{Maximum (left graph) and mean (right) accuracy of classifiers as a function of a number of variables (dimensions); standard deviations were not plotted for the sake of readability, they are of the order of $0.02$}
\label{fig:te1}
\end{figure}

In Fig.~\ref{fig:te1} the best (left) and the mean (right) accuracy of threshold forecasting (see subsection~\ref{bp2}) classifiers with a given dimensionality is presented. The best results are achieved for support vector machines with a radial-basis core function. These classifiers achieved an accuracy of 96\%. Also in this case adding dimensions does not necessarily increase the quality of a classifier; for SVM with a tangential core adding new dimensions decreases the accuracy of the best classifier. Again, the mean accuracy of the classifier tends to increase with every additional dimension, however.

\begin{figure}[htb]
\centerline{
\includegraphics[width=12cm]{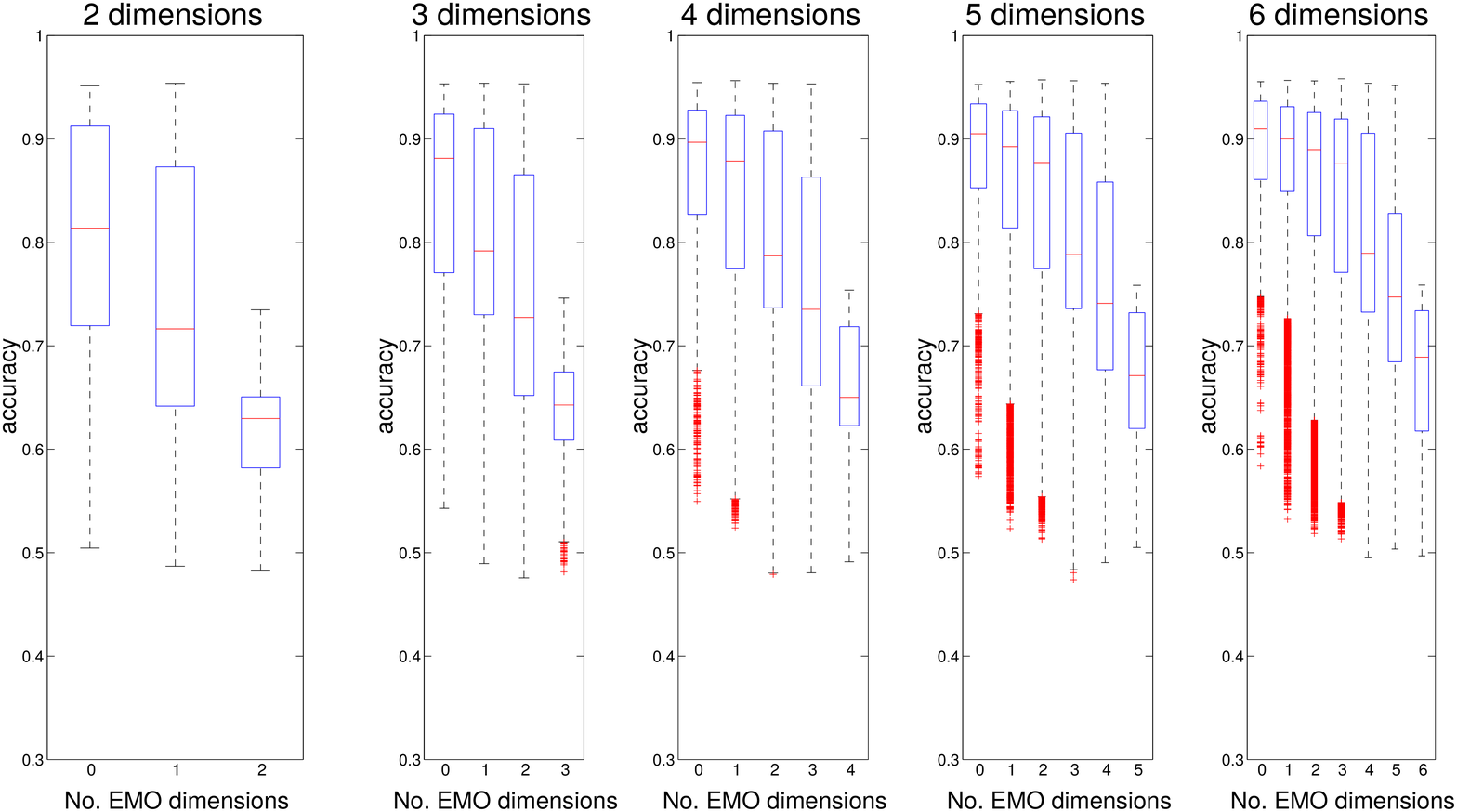}}
\caption{Box plot of \textbf{accuracy} for different numbers of variables used by \textbf{various classifiers} as a function of number of emotion-related variables;}
\label{fig:te2}
\end{figure}

\begin{figure}[htb]
\centerline{
\includegraphics[width=12cm]{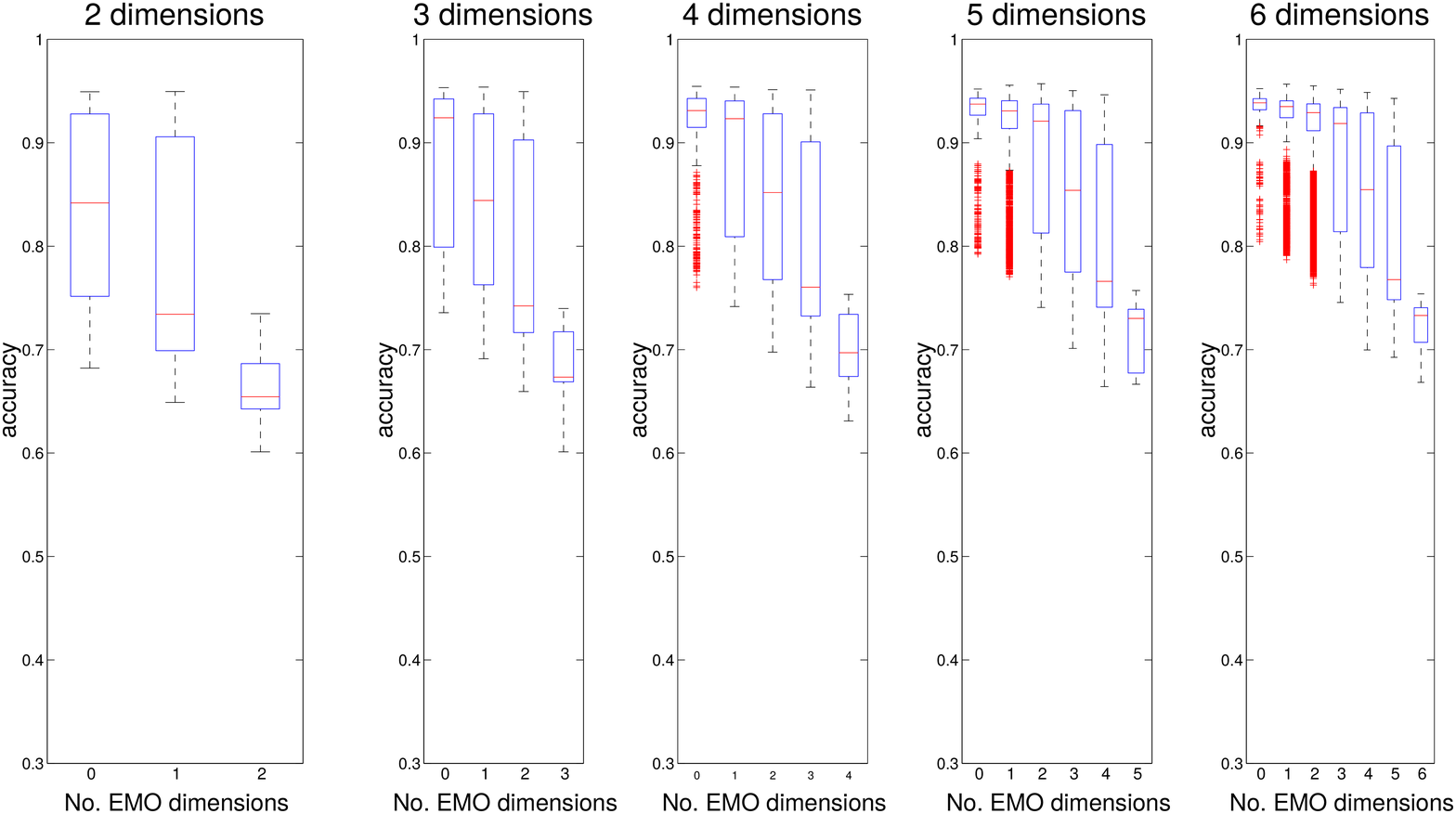}}
\caption{Box plot of \textbf{accuracy} for different numbers of variables used by \textbf{SVM with a radial basis core} as a function of number of emotion-related variables;}
\label{fig:te3}
\end{figure}

Fig.~\ref{fig:te2} shows the relationship between the accuracy and the number of emotional dimensions for all methods, while in Fig.~\ref{fig:te3} -- for support vector machines with a radial-basis core (the most accurate classifier for the problem). Also in this case, using emotional variables provided no classifier enhancing information.

\subsection{Peak classification}
\begin{figure}[htb]
\centerline{
\includegraphics[width=12cm]{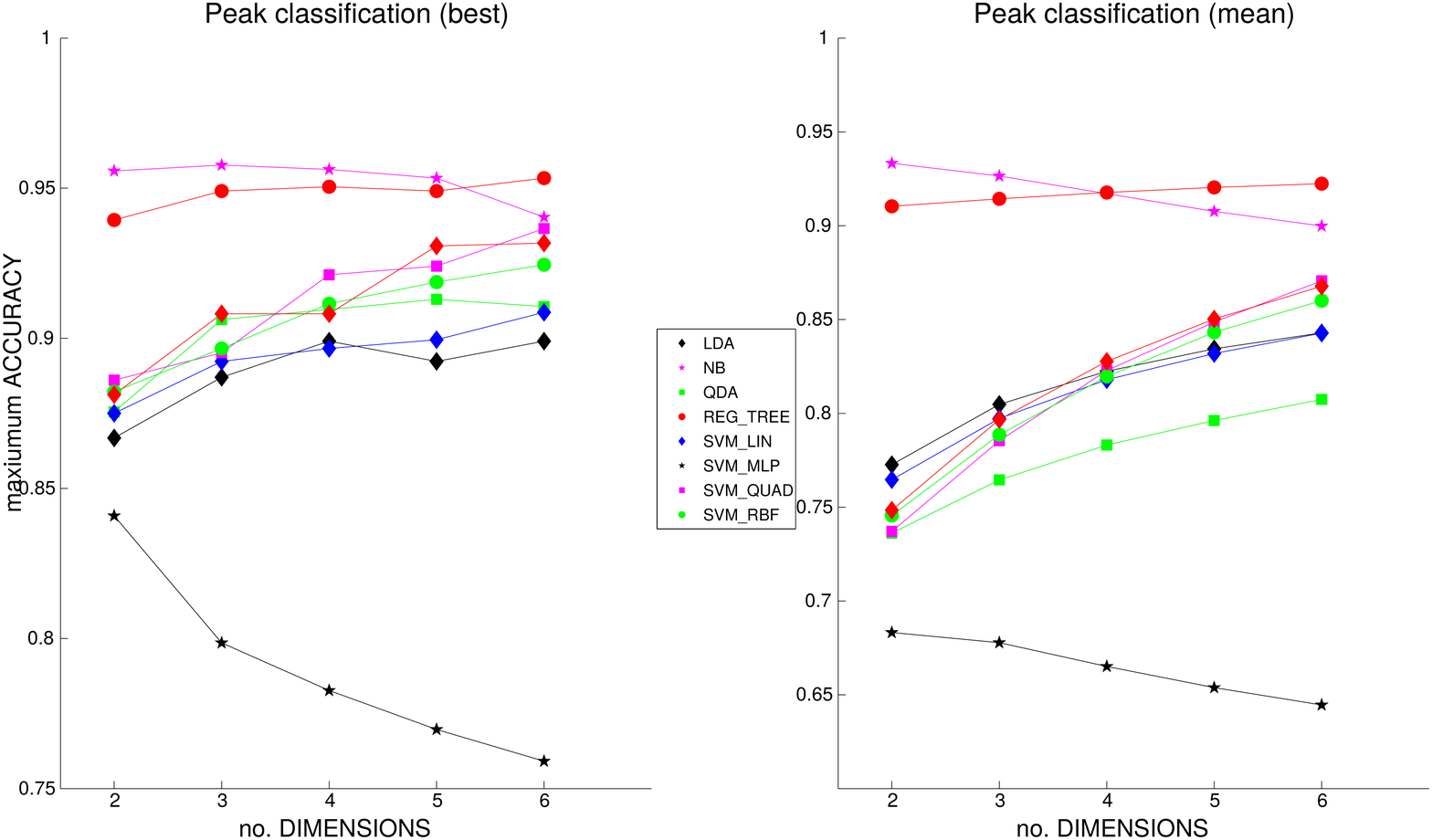}}
\caption{Maximum (left graph) and mean (right) accuracy of classifiers as a function of a number of variables (dimensions); standard deviations were not plotted for the sake of readability, they are of the order of $0.02$}
\label{fig:pc1}
\end{figure}

The third problem is fundamentally different from the previous ones. In Fig.~\ref{fig:pc1} the best (left) and the mean (right) accuracy of  the peak type classifiers (see subsection~\ref{bp3}) with a given dimensionality is presented. The results are a little surprising – the best accuracy (about 96\%) was achieved for the naive Bayes classifier with 3 features. Similar results were achieved by naive Bayesian classifiers with a different number of dimensions and a regression tree. The mean accuracy of the classifier tends to increase with every additional dimension for most of the methods, but not for the best and the worst method.

\begin{figure}[htb]
\centerline{
\includegraphics[width=12cm]{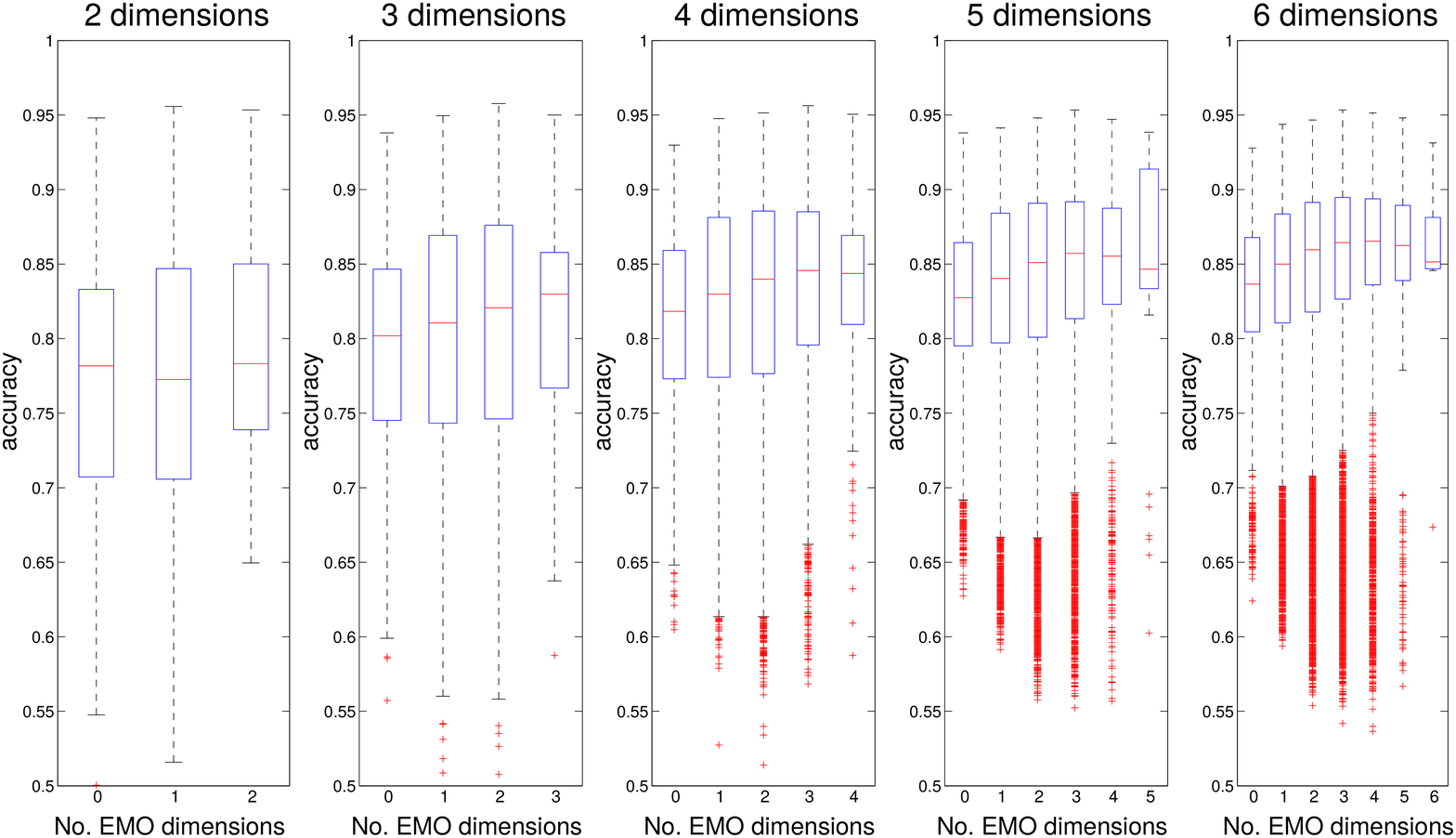}}
\caption{Box plot of \textbf{accuracy} for different numbers of variables used by \textbf{various classifiers} as a function of number of emotion-related variables;}
\label{fig:pc2}
\end{figure}

\begin{figure}[htb]
\centerline{
\includegraphics[width=12cm]{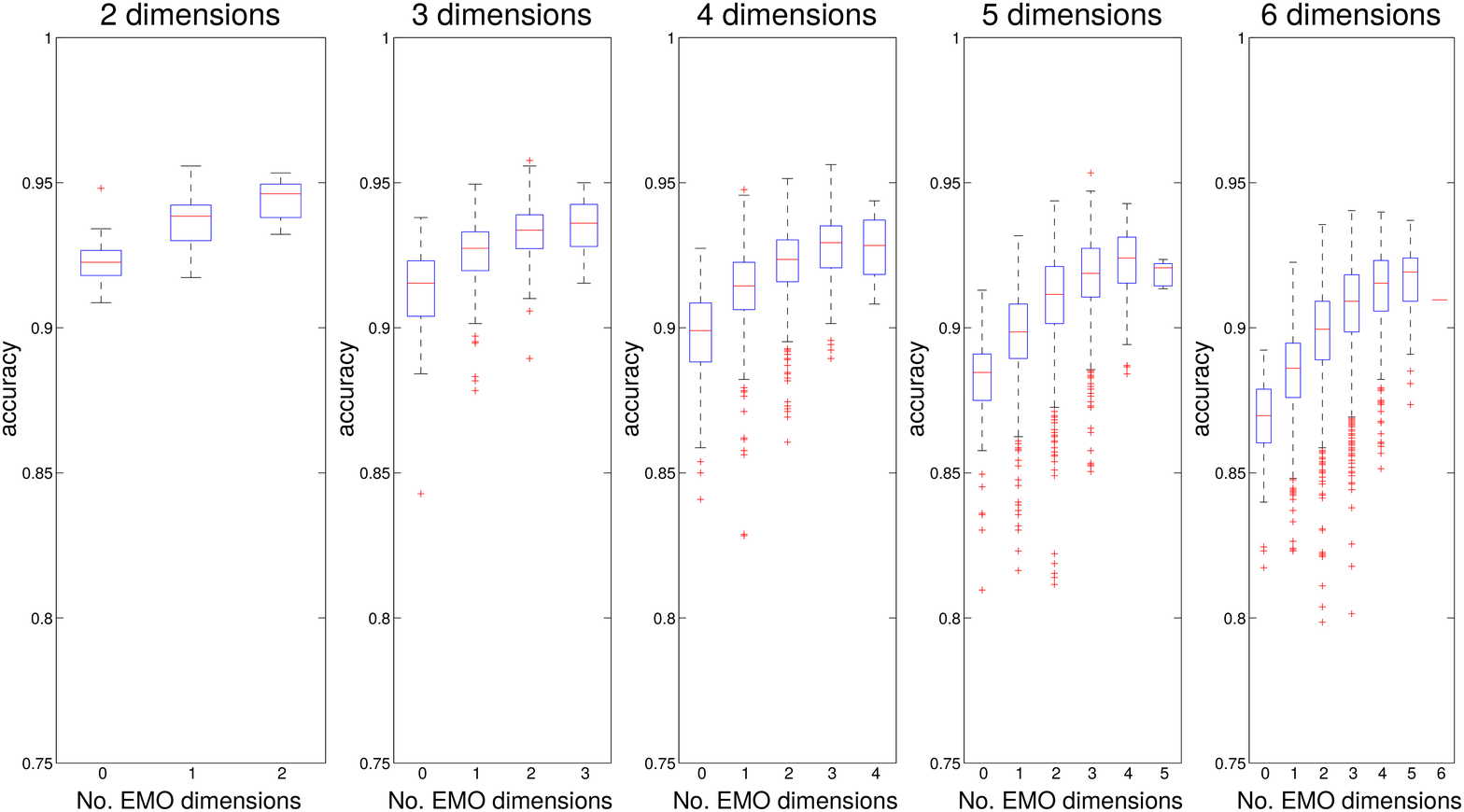}}
\caption{Box plot of \textbf{accuracy} for different numbers of variables used by \textbf{naive Bayes classifier} as a function of number of emotion-related variables;}
\label{fig:pc3}
\end{figure}

Fig.~\ref{fig:pc2} shows the relationship between the accuracy and the number of emotional dimensions for all methods, while in Fig.~\ref{fig:pc3} -- for naive Bayes classifiers (the most accurate classifiers for the problem). In this case, using emotional dimensions clearly increases the accuracy of the classifiers.

\section{Conclusions} 
This research has been conducted on Twitter activity and sentiment data. We investigated the role of emotional dimensions in classification but our approach could be easily adapted to any classification (or prediction) study of certain systems. To achieve this, change the dataset to another aggregated macroscopic collection of data from a system with unknown internal microscopic dynamics and emotional dimensions to problem-specific variables. The insights provided may confirm the usefulness of the variables and identify the best suited methods and variable sets for a specific problem.

The results show the significance of the emotional content of Twitter posts in the peak classification study related to special events (i.e. medal winning), but employing emotional dimensions in predicting activity brings no additional insights.

A set of popular data mining techniques' performance was analyzed for activity prediction and peak classification. The results are interesting because of evidence about the importance of emotions in social media and a wide spectrum of applications (e.g. predicting time series, classifying the origins of peaks).

The relationship between the accuracies of a set of basic data mining methods and number of features as well as between their accuracy and the number of emotional features was also analyzed. The benchmark problems consisted of (A) trend prediction, (B) forecasting of threshold crossing and (C) peak classification.

Depending on the problem, the best classifiers had an accuracy of (A) 66\%, (B) 96\% and (C) 96\%.

Adding a dimension to the benchmark problems increased the mean classifier accuracy for most cases but the best classifier for a given method is not necessarily the one with the largest number of dimensions. In a few cases, adding dimensions decreases the best accuracy and the mean classifier accuracy.

Using an emotional dimension has no impact for (A) and (B), while clearly enhances accuracy for (C).

The results confirm the usefulness of classifiers, give insights about applying additional features and dimensions and show that the methods analyzed can be applied to quantitative descriptions of social media phenomena.

\section{Acknowledgements}
The research leading to these results has received funding from the EU Seventh Framework Programme (FP7/2007-2013) under grant agreement no. 231323 (\textit{Collective Emotions in Cyberspace} project -- CyberEMOTIONS).
J.A.H. has also been partially supported by the Russian Scientific Foundation, proposal \#14-21-00137. 
J.S. acknowledges support of the European Union in the framework of the European Social Fund through the Warsaw University of Technology Development Programme, realized by the Center of Advanced Studies. We are also thankful to Tomasz Ryczkowski for Twitter data pre-processing and medal data gathering.

\bibliographystyle{unsrtmy}
\bibliography{APPA_Choloniewski_Sienkiewicz_Holyst_v5}

\clearpage
\begin{table}[]
\section*{Appendix A -- Events with British medalists}
\centering\noindent\makebox[\textwidth]{
\small
\begin{tabular}{|l|l|l|l|l|l|}
\hline
\textbf{No.} & \textbf{Date} & \textbf{Time} & \textbf{Sport} & \textbf{Event} & \textbf{Medal} \\
\hline\hline
1&29 July&18:45&Cycling&Women's road race&Silver\\ \hline
2&29 July&23:45&Swimming&Women's 400 m freestyle&Bronze\\ \hline
3&30 July&22:00&Gymnastics&Men's artistic team all-around&Bronze\\ \hline
4&31 July&16:00&Equestrian&Team eventing&Silver\\ \hline
5&1 August&15:00&Rowing&Women's coxless pair&Gold\\ \hline
6&1 August&15:30&Rowing&Men's eight&Bronze\\ \hline
7&1 August&19:15&Cycling&Men's time trial&Bronze\\ \hline
8&1 August&19:15&Cycling&Men's time trial&Gold\\ \hline
9&2 August&18:45&Canoeing&Men's slalom C-2&Gold\\ \hline
10&2 August&18:45&Shooting&Men's double trap&Gold\\ \hline
11&2 August&18:45&Canoeing&Men's slalom C-2&Silver\\ \hline
12&2 August&19:00&Judo&Women's 78 kg&Silver\\ \hline
13&2 August&21:15&Cycling&Men's team sprint&Gold\\ \hline
14&3 August&15:15&Rowing&Men's single sculls&Bronze\\ \hline
15&3 August&15:15&Rowing&Women's double sculls&Gold\\ \hline
16&3 August&15:45&Rowing&Men's coxless pair&Bronze\\ \hline
17&3 August&21:45&Cycling&Men's team pursuit&Gold\\ \hline
18&3 August&22:15&Cycling&Women's keirin&Gold\\ \hline
19&4 August&14:45&Rowing&Men's coxless four&Gold\\ \hline
20&4 August&15:00&Rowing&Women's lightweight double sculls&Gold\\ \hline
21&4 August&21:15&Cycling&Women's team pursuit&Gold\\ \hline
22&4 August&00:00&Athletics&Women's heptathlon&Gold\\ \hline
23&4 August&00:30&Athletics&Men's long jump&Gold\\ \hline
24&4 August&01:00&Athletics&Men's 10,000 m&Gold\\ \hline
25&5 August&17:45&Sailing&Finn class&Gold\\ \hline
26&5 August&19:15&Gymnastics&Men's pommel horse&Bronze\\ \hline
27&5 August&19:15&Gymnastics&Men's pommel horse&Silver\\ \hline
28&5 August&19:45&Tennis&Men's singles&Gold\\ \hline
29&5 August&20:30&Tennis&Mixed doubles&Silver\\ \hline
30&5 August&22:00&Cycling&Men's omnium&Bronze\\ \hline
31&5 August&00:15&Athletics&Women's 400 m&Silver\\ \hline
32&6 August&17:45&Gymnastics&Women's uneven bars&Bronze\\ \hline
33&6 August&20:00&Equestrian&Team jumping&Gold\\ \hline
34&6 August&21:15&Cycling&Men's sprint&Gold\\ \hline
35&7 August&16:15&Sailing&Men's sailboard&Silver\\ \hline
36&7 August&17:15&Triathlon&Men's triathlon&Bronze\\ \hline
37&7 August&17:15&Triathlon&Men's triathlon&Gold\\ \hline
38&7 August&19:15&Equestrian&Team dressage&Gold\\ \hline
39&7 August&20:15&Cycling&Women's omnium&Gold\\ \hline
40&7 August&21:15&Cycling&Men's keirin&Gold\\ \hline
41&7 August&21:30&Cycling&Women's sprint&Silver\\ \hline
42&7 August&00:00&Athletics&Men's high jump&Bronze\\ \hline
43&9 August&19:00&Equestrian&Individual dressage&Bronze\\ \hline
44&9 August&19:00&Equestrian&Individual dressage&Gold\\ \hline
45&9 August&19:45&Boxing&Women's flyweight&Gold\\ \hline
46&9 August&01:30&Taekwondo&Women's 57 kg&Gold\\ \hline
47&10 August&16:45&Sailing&Women's 470 class&Silver\\ \hline
48&11 August&12:45&Canoeing&Men's K-1 200 m&Gold\\ \hline
49&11 August&23:00&Athletics&Men's 5,000 m&Gold\\ \hline
50&11 August&00:15&Boxing&Men's bantamweight&Gold\\ \hline
51&11 August&01:15&Diving&Men's 10 m platform&Bronze\\ \hline
52&12 August&18:45&Boxing&Men's super heavyweight&Gold\\ \hline
\end{tabular}}
\label{tab_medals}
\end{table}
\end{document}